\documentclass{article}
\pdfoutput=1
\usepackage{booktabs} 

\usepackage{caption}
\usepackage{subcaption}
\usepackage{makecell}
\usepackage{hyperref}
\usepackage{amssymb}
\usepackage{amsmath}
\usepackage{graphicx}
\usepackage{natbib}
\usepackage{authblk}

\begin{document}
\title{Inter-Session Modeling for Session-Based Recommendation\footnote{This work was carried out at the Telenor-NTNU AI-Lab, hosted by the Department of Computer Science, Norwegian University of Science and Technology.}}

\author[1,2]{Massimiliano Ruocco}
\author[1]{Ole Steinar Lillest{\o}l Skrede}
\author[1]{Helge Langseth}
\affil[1]{Department of Computer Science, Norwegian University of Science and Technology}
\affil[2]{Telenor Research}

\maketitle

\begin{abstract}
In recent years, research has been done on applying Recurrent Neural Networks (RNNs) as recommender systems. Results have been promising, especially in the session-based setting where RNNs have been shown to outperform state-of-the-art models. In many of these experiments, the RNN could potentially improve the recommendations by utilizing information about the user's past sessions, in addition to its own interactions in the current session. A problem for session-based recommendation, is how to produce accurate recommendations at the start of a session, before the system has learned much about the user's current interests.
We propose a novel approach that extends a RNN recommender to be able to process the user's recent sessions, in order to improve recommendations. This is done by using a second RNN to learn from recent sessions, and predict the user's interest in the current session. By feeding this information to the original RNN, it is able to improve its recommendations.
Our experiments on two different datasets show that the proposed approach can significantly improve recommendations throughout the sessions, compared to a single RNN working only on the current session. The proposed model especially improves recommendations at the start of sessions, and is therefore able to deal with the cold start problem within sessions.
\end{abstract}

\section{Introduction}
\label{sec:introduction}
In a session-based setting, the actions of the user within a session are correlated. This means that a recommender system can observe the user's actions and improve the recommendations as the system learns more about the user's interest. Recently, RNNs has been shown to work well in the session-based setting \cite{HidasiKBT15,Twardowski2016,Zhang2014,LiuWWL016}. RNNs are naturally good at working with sequences of data, because they have an internal memory storing the past observations, and the ability to update and discard information in their memory. Therefore, a RNN will make more accurate recommendations as it learns more about a user. This also means that a simple RNN will struggle to make good recommendations at the start of a session. The advantage of employing an RNN over many other recommendation/prediction models, is that it naturally considers the order of sequences. Many other models use the relaxed assumption that the order does not matter. Solutions that take the sequence into account are possible, but RNNs considers the order of sequences in a very natural way that few other models do.

In many of the services where user interaction is session-based, the users are logged in and their actions can be stored. If this is done, a recommender system can get access to the user's history, which it can use to improve the recommendations. A RNN could possibly use information from a users history to make precise recommendations from the start of a new session, and possibly improve all recommendations in that session.
In the session-based scenario, the user history consists of a ordered sequence of sessions, and the RNN could also be used to process user history, as we show in our proposed architecture.

Session-based recommender systems that only consider the current session face the task of doing recommendations based only on small set of interactions. Collaborative filtering approaches fall short here, and usually content-based filtering is used instead, i.e. recommending similar items. The data in a session consists of a sequence of user actions. The sequences may vary in length. Also, the actions within a session are likely to be dependent. These properties fit well with RNNs, and therefore they can perform well in this setting.
Intuitively, a RNN should be able to capture dependencies between items, like content-based filtering, but with its memory capabilities a RNN should also be able to consider the whole session, which could lead to more accurate predictions. Recent papers have shown promising results in using RNNs for session-based recommendation \cite{HidasiKBT15, TanXL16}.

When user history is available to the recommender system, collaborative filtering approaches, such as matrix factorization, can perform well. A RNN can still perform well, but it will probably struggle at the start of the sessions until it has learned what the user is interested in. This problem could of course be fixed if the RNN was able to learn from the user history before starting the session, and thus have a foundation to make recommendations on, right from the start. This extra information could potentially improve the overall recommendations, and especially the initial recommendations in each session.
In a session-based setting where user history is available, the user history consists of past sessions. So the user history is a sequence of sessions, and each session is a sequence of events. This brings us back to our motivation for this work. The idea is to use a RNN to do predictions within a session, and to employ another RNN layer to contribute in the prediction of the users interests for the next session. Let's consider the example of a user shopping on a e-commerce site. One day he might buy a laptop, some days later he will buy some hiking gear, and some days after that he buys accessories to the laptop he bought in the first session. This illustrates that the users interest in a session can be dependent on what he did in earlier sessions, and that just considering the previous session is not enough.

In this work we will investigate how a RNN can be used to learn from user histories, and thereby improve the straightforward use of RNN in session-based recommendations. In particular, the contributions will be the following: (1) We introduce the concept of inter session learning, (2) We propose a way to learn the inter session behavior together with the intra-session one (3) We validate the effectiveness of the model by an extensive set of experiments, (4) We show the effectiveness of the method in tackling the cold start problem.

\section{Related Work}
\label{sec:relatedwork}
In recent year, different deep learning techniques have been successfully employed in the recommendation context. In particular the use of RNN has been shown to be promising in the area of session-based recommendation. In this section we will present the state of the art in these areas.
The idea of using RNN in a straightforward way for Session-based recommendation has been first introduced in \cite{HidasiKBT15}. Further works, extending this initial work still employing an RNN for the recommendation task, have been presented in \cite{Tan2016IRN, Hidasi2016PRN, Jing2017NSR, Wu2017RRN, Song2016MDL,LiuWWL016, Twardowski2016}. 
In \cite{HidasiKBT15} the authors shows that a basic RNN for session-based recommendations, can achieve remarkable results. They also deal with sparsity issues, and introduce a new ranking loss function for training the network. They experimented with two different datasets. Both datasets contain sequences of user clicks with timestamps. One dataset has clicks on items from an e-commerce site from the RecSys Challenge 2015\footnote{Recsys challange 2015: \url{http://2015.recsyschallenge.com/challenge.html}}, while the other contains clicks on videos from a YouTube-like platform. Various modifications of the network were tested. 
In \cite{Tan2016IRN} the authors explored various ways for improving the model proposed in \cite{HidasiKBT15}, used as baseline. They experiment with techniques that have worked well when neural networks have been applied to other problems, to see if those techniques can improve performance of a RNN session-based recommender as well. They experimented their proposed models on the same dataset based on four different way of improving the original model. They improved the original methods by mainly a) applying a data augmentation technique for tackling with the problem of session varying in length, b) accounting for temporal shifts in the data distribution related to users behavior to tackle cases in which the products are released in different periods of time. 
So far, the presented works refer to models that performs predictions solely based on the items clicked, where the items are only represented by an ID often in the form of a one-hot vector. Clearly, additional information, both about the item and about the sessions, could help improve the predictions. Some possible additional information about the item is be the category of the item, an image, and a textual description of the item. Additional information about the session could be timestamps of the clicks, geo-location of the user, and weather. In \cite{LiuWWL016}, for example, the authors suggest that modeling the time of a session and the transition time between events in the session both can give better performance. Assumption about the temporal dimension is also done in other works. In \cite{Song2016MDL}, the main assumption of the authors is that user interests change over time. As an example, in \cite{Elkahky2015MDL}, it was shown that users who visited \url{spiegel.de}, a popular German news portal, were likely to be interested in football related news. The reason was that the data was collected around the time of the Football World Cup of 2014. Similarly, user interests may change over time, e.g. during Summer and Christmas. The authors propose to use a model that combines static and temporal user features. The static features are learned by using the full training set, while the temporal features are learned by only training on the most recent examples. The time aspect is also considered in \cite{Jing2017NSR} for the task of  "Just in Time" recommendation, where the objective is to recommend the right items at the right time. The inter-session dependency are here considered to learn recurrent user activities by a LSTM-based architecture. The work proposed in \cite{Wu2017RRN} start from the assumption that many of the current state of the art approaches and methods for recommendation are lacking when it comes to temporal and causal aspects inherent in the data. In particular they state that user profiles and movie attributes are generally considered static.They proposed a RNN-based model considering these aspects and modelling the user and movie dynamics. The model is shown to be able to capture temporal patterns in rating data, outperforming previous works in term of prediction accuracy 	
Often, other extra information are available. In \cite{Twardowski2016}, the authors propose a model employing the item extra information available for example in some e-commerce site such as the type of action the user performed (i.e., viewing an item, adding it to the basket, removing it from the basket, or buying it). They propose a RNN-based model that makes use of this information for recommendations. The proposed model sends embedded event information through a RNN layer, the output is concatenated with an embedded item representation, before being sent through feed forward layers to produce a prediction. On a dataset with rich search contextual information, the proposed RNN model performs significantly better than other compared models and baselines. While on a dataset with less events and data, the RNN-based model performed worse than a matrix factorization model that was also customized to utilize event information. Sometimes, in some e-commerce site, items are also described and associated with information such as picture and textual description. In \cite{Hidasi2016PRN}, the authors explore the possibility of employing this richer features representation in a number parallel RNN architectures to model sessions on the clicks and the rich features (text and images) of the clicked item. 
\section{The II-RNN Architecture}
\label{sec:iirnn_architecture}

In the session-based setting, the user's actions might depend on all earlier actions in the session, not just the previous one. How the dependencies between the actions work, will vary between different domains. For example, on a news site, if a user reads articles about German news and international sports, that user will probably be interested in reading news articles about German sport, while for a online grocery shopping site, past actions might indicate that the user will not be interested in similar items. If the user has added bread and milk to his basket, he will probably not add anymore bread or milk to that basket. But if the user has only added milk to the basket, it might be interested in adding bread as well. 

\subsection{Main Idea}

RNNs work well in the session-based recommendation scenario because it can process sequences of user actions, and create an internal representation of the user's interests. Also, it does not assume that all actions indicate interest in something, it can learn to interpret actions as sign of disinterest. 
As discussed in the previous Section, the RNN model achieves state-of-the-art performance on session-based recommendation problems.

In addition to the short term dependencies between actions within a session, there are usually long term dependencies between actions from different sessions. E.g., a user that was interested in news articles about golf in his previous session(s), will probably also have that interest in his current session. Or a user that bought a new laptop in a recent previous session, will probably not be interested in buying another one in the current session, but he might be interested in accessories to the laptop he bought. This means that it should be possible to improve the recommendations for a session-based recommender system, by giving it information about the user's interaction history. Furthermore, one of the reasons that a RNN works well for recommendations within a session, is that it is able to process the sequence of the session events. Similarly, we believe that the order of the sequence of earlier sessions can be important. En example could be a person that regularly does his grocery shopping online. If he buys bread in one session, then he will probably not be interested in buying another one within the next few sessions. On the other hand, he is probably going to buy bread soon if  he has not done so during the last few sessions.

Since RNNs work well for recommendations on sequences of events within a session, and because the sessions themselves form a sequence, we think that a RNN could work well to process the sequence of sessions as well.

The main idea is to use one RNN to process the events within a session, as has been done before, and to enhance the recommendations from this by using a second RNN to process a user's recent sessions and help the first RNN with a initial prediction about the current session. In other words, a RNN that works on a inter-session level, provides the initial hidden state for a RNN that works on a intra-session level. We will refer to this model as II-RNN (Inter-Intra RNN).

\subsection{Problem Formulation}
In the session-based recommendation scenario, there is a system with a set of items that a user can interact with; note that the term "item" is used in a broad sense here. We experiment with the proposed models using two different datasets, where the possible recommendations are sub-forums of a discussion site and artists on a music website, respectively. The datasets are described in Section \ref{sec:experimentalsettings}.

Let $N$ be the set of items in the system, and $n_{v} \in \mathbb{R}^{d}$ is the embedded representation of item $v$. Each user $u$ has a interaction history $S^{u} = \{S^{u}_{t_{1}} , S^{u}_{t_{2}} , \ldots\}$, where $S^{u}$ is a session of interaction by user $u$ at time $t_{i}$. The session history is ordered temporally by $t_{i}$. The session length is $|S_{u}|$. Each session $S^{u}_{t_{i}}$ consists of a collection of events $\{{e}^{u}_{t_{i},j} \in \mathbb{R}^{m} | j = 1, 2, ..., |S_{t_{i},u}|\}$, where ${e}^{u}_{t_{i},j}$ is the representation of event $j$ in the session. 
While events can be any type of interaction in general, events in this work will simply be items the user interacts with. Hence, an event will relate directly to an item $v$. All recommendation models we experiment with use an item id, $i_{v} \in \{1,2,...,|N|\}$, as input for each item. However, the RNN models retrieves the corresponding embedded representation $\mathbf{n}_{v}$ for each $i_{v}$, and feed those into the RNN layer of the model.
The common task for all the recommendation models we experiment with is to predict each consecutive item in a session $S^{u}_{t_i}$. That is, for a sub-session  $\{e^{u}_{t_{i}, 1}, e^{u}_{t_{i}, 2},\ldots, e^{u}_{t_{i}, j}\}$ of $S^u_{t_i}$, the system is to predict $e^{u}_{t_{i}, j+1}$. This is repeated for $j = 1, 2, \ldots |S^{u}_{t_{i}}|-1$.
A recommendation $R_{j}$ is an ordered list of $k$ recommended items, where we would want to see the next item, $e^{u}_{t_{i}, j+1}$, as close to the top as possible. 

\subsection{Model Description}
II-RNN combines  the modeling of the inter-session with the intra-session behavior of a single architecture. 
The first the model is similar to the one used in \cite{HidasiKBT15}, and the model proposed in \cite{HidasiKBT15} will therefore serve as a baseline to compare the II-RNN model to.

\begin{figure}
  \centerline{\includegraphics[width=0.6\textwidth]{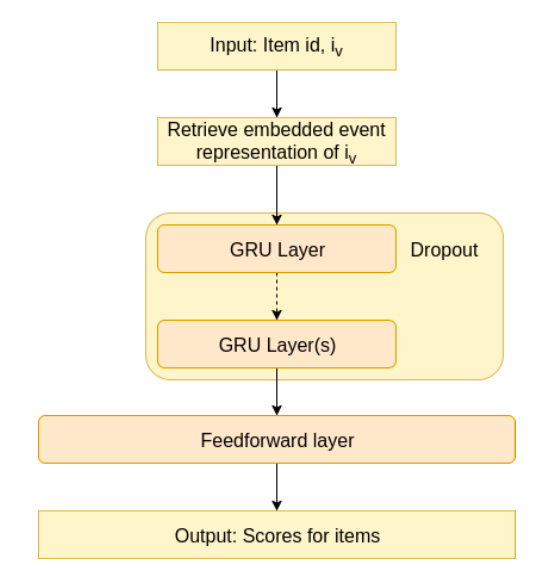}}
  \caption{The intra-session RNN}
  \label{fig:intraRNN}
\end{figure}

\paragraph{Intra-session RNN}
The intra-session RNN produces recommendations by processing the sequence of items in a session. Figure \ref{fig:intraRNN} illustrates the model. This model is very similar to the one in \cite{HidasiKBT15} and other papers. We do not use one-hot encodings as input, but use item embeddings directly. Mathematically these two methods are equivalent, but in practice this saves us the computation required to create the one-hot vectors. When the set of items is huge, creating a mini-batch of one-hot vectors will require a large amount of memory, which can be a problem.

The embedded item representation is sent through one or multiple layers of GRU, and dropout is applied to these layers. Afterwards a feed-forward layer is used to scale up the vector to $\mathbb{R}^{|N|}$. The output vector is then $[o_{v_{1}} o_{v_{2}} ... o_{v_{|N|}}]$ where $o_{v_{i}}$ is a score for item $v_{i} \in N$. The list of recommendations, $R_{j}$ is then created by taking items corresponding to the $k$ highest scores, sorted by their score.
Training is done with the Adam algorithm for stochastic gradient descent \cite{KingmaB14}, and the loss is calculated with cross entropy. The target output is a score of $0$ for all items, except for the relevant item which should get a score of $1$. This means that we treat the recommendation problem as a classification problem. That is, given the users recent activity, predict the next item he will interact with. This works because the model predicts scores for how likely it believes that each item is the correct class, and these scores then form a natural way of ranking the recommendations.

\begin{figure*}	
	\centering
	\begin{subfigure}[t]{3.9in}
		\centering
        \includegraphics[width=1.0\textwidth]{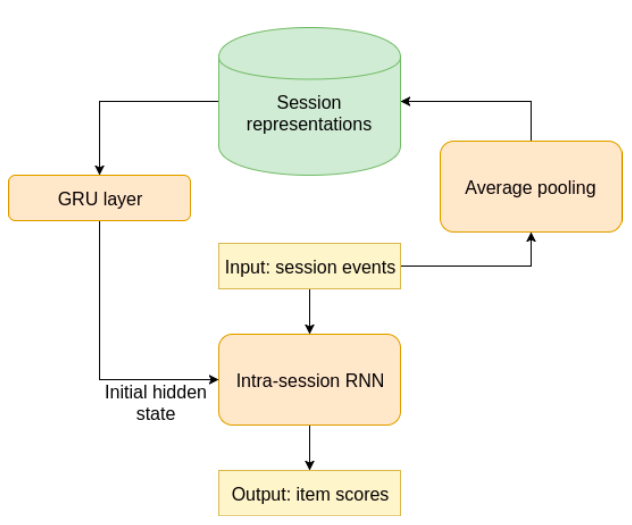}
		\caption{The II-RNN Model with average pooling to create session representations from items.}\label{fig:iiRNNavgpooling-used}
	\end{subfigure}
    \hspace{0.1\textwidth}
	\begin{subfigure}[t]{3.9in}
		\centering
        \includegraphics[width=0.85\textwidth]{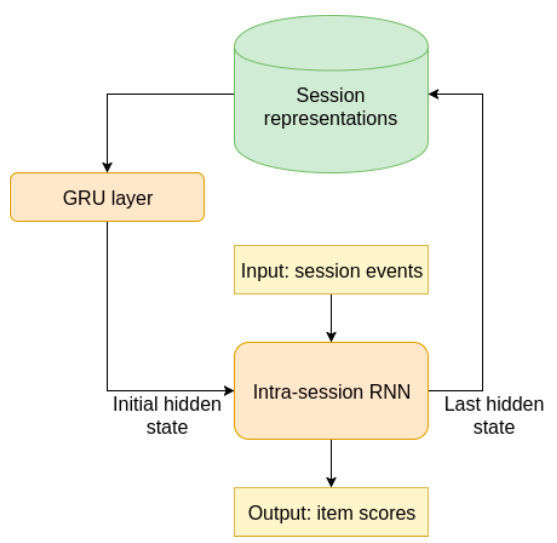}
		\caption{The II-RNN Model where the last hidden state of the intra- session RNN is stored as the session representation.}\label{fig:iiRNNlasthidden}
	\end{subfigure}
	\caption{The proposed II-RNN architectures}\label{fig:iiRNN}
\end{figure*}

\paragraph{II-RNN Model}
Although the intra-session RNN can achieve a strong performance, it starts out in each session without any knowledge about the user. It learns about the user's interests throughout the session, but all that information is discarded again at the end of that session. The II-RNN can improve upon the intra-session RNN, because it takes the user's previous sessions into account, and supplies the intra-session part with information at the start of each new session. Figure \ref{fig:iiRNN} illustrates the II-RNN.
For each session $S^{u}_{t_{i}}$ in a user's interaction history  $S_{u}$, let $\mathbf{s}^{u}_{t_{i}}$ be an embedded vector representation of that session. The input to the inter-session RNN layer (the GRU layer in Figure \ref{fig:iiRNN}) is then the sequence $\{{\mathbf{s}}^{u}_{t_{z-g}}, {\mathbf{s}}^{u}_{t_{z-g+1}}, \ldots, {\mathbf{s}}^{u}_{t_{z}}\}$, where ${\mathbf{s}}^{u}_{t_{z}}$ is the representation of the most recent session, and $g$ is the number of recent sessions that should be processed. The initial hidden state, $H_{0}$, of the intra-session RNN is then set to final output of the inter-session RNN. In other words, the inter-session RNN produce the initial hidden state of the intra-session RNN, based on a series of vector representations of the most recent sessions for the given user. The output of the inter-session RNN is calculated before the intra-session RNN starts producing predictions.

We apply two different methods of producing the session representations ${\mathbf{s}}^{u}_{t_{i}}$. One is the average of the embedded vector representations of the items in the session, as illustrated in Figure \ref{fig:iiRNNavgpooling-used}. The other is to simply use the the last hidden state of the intra-session RNN as the session representation, illustrated in Figure \ref{fig:iiRNNlasthidden}. Even though the final hidden state can contain more useful information learned by the intra-session RNN, it is more a representation of the end of the session, rather than the whole session. Since the hidden state is produced by a RNN, it will depend on the order of the sequence of items in a session, while the average of the embeddings is unaffected by the order of the items.

%

\section{Experimental Setting}
\label{sec:experimentalsettings}

\subsection{Datasets}
We experimented with two different datasets: The first is a dataset on user activity on the social news aggregation and discussion website Reddit\footnote{Subreddit interactions dataset: \url{ https://www.kaggle.com/colemaclean/subreddit-interactions}}. This dataset contains tuples of usernames, a subreddit where the user made a comment to a thread, and a timestamp for the interaction. The second dataset contains listening habits of users on the music website Last.fm \cite{Bertin-Mahieux2011}. This dataset contains tuples of user, timestamp, artist, and song listened to.

\paragraph{Reddit dataset}
The Reddit dataset contains a log of user interaction on different subreddits (sub-forums), with timestamps. Here, an interaction is when a user adds a comment to a thread. Since the dataset itself, does not split the interactions into sessions, we did this manually when preprocessing the dataset. To do this we analyzed the dataset and specified a time limit for inactivity. Using the timestamps, we let consecutive actions that happened within the time limit belong to the same session. That is, for a specified time limit $\Delta_t$, and a list of a user's interactions $\{a_{t_{0}} , a_{t_{1}} , \ldots, a_{t_{n}}\}$, ordered by their timestamps $t_i$, two consecutive interactions $a_{t_{i}}$ and $a_{t_{i+1}}$ belong to the same session if and only if $t_{i+1} \leq t_{i} + \Delta_t$. We set the time limit to 1 hour ($3600$ seconds).
Note that users, in addition to commenting on threads, also do browsing and reading. Therefore it makes sense to set a time limit that allows for some time between the interactions captured in the dataset. Also some users are more active than others, some users are mostly passive consumers who rarely comments. So, it is impossible to set a time limit that fits all users. However, it is important that the time limit is large enough that the average session contains a fair amount of interactions, but small enough so that it is reasonable to assume that the interactions are dependent on each other.

\paragraph{Last.fm dataset}
We also had to split each user's history into sessions manually for the Last.fm dataset. We used the same approach as for the Reddit dataset, but here we used $30$ minutes ($1800$ seconds) as the time limit. Also, we faced the problem that the dataset contains an overwhelming amount of songs. Since our recommendation models produce a score for each possible item, the huge amount of songs caused a memory requirement problem. To solve this, we simplified the dataset by ignoring the specific song of each user interaction and only use the artists. This reduce the item set to a manageable size.

\subsection{Preprocessing}
After the initial manual splitting into sessions, we used the same preprocessing for all three datasets. In the Reddit and Last.fm datasets, there were many items that repeated consecutively. We are not interested in a recommender system that learns to predict the last seen item, therefore we removed all consecutively repeating items, and only kept one instance. Furthermore, the RNN models need to have a specified maximum length of the sessions, because they must be unrolled in order to be trained. To deal with this, we set the maximum length, $L$, of a session to $L=20$. Sessions that had a length $l$ of $L < l < 2L$ were split into two sessions. This was done because we did not want to throw away all sessions that were too long, but splitting very long sessions create many sessions that should not be separate sessions, since the events in them depend on each other. However, there were some unreasonable long sessions that probably originate from bots or some other error source. These were removed with the $2L$ limit for session lengths. With this scheme, the majority of the sessions from all datasets were kept.
Sessions of length $l < 2$ were removed, and users with less than $3$ sessions were also removed. Finally, the datasets were split into a training set and a test set on a per user basis. For each user, $80\%$ of his sessions were placed in the training set, and the remaining in the test set. Each user's sessions were sorted by the timestamp of the earliest event in the session, and the test set contains the most recent sessions of each user.
Table \ref{tab:preproc} shows statistics for the two datasets after preprocessing (before splitting into training and test sets).

\begin{table}
\begin{center}
  \begin{tabular}{lcc}
    \toprule
     &Reddit&Last.fm\\
    \midrule
    Number of users & 18.271& 977\\
    Number of sessions & 1.135.488& 630.774\\
    Sessions per user & 62,1& 645,6\\
    Average session length & 3,0& 8,1\\
    Number of items & 27.452& 94.284\\
  \bottomrule
\end{tabular}
\end{center}
\caption{Statistics for the datasets after preprocessing}
\label{tab:preproc}
\end{table}

\subsection{Baselines}
In addition to the following baselines, the intra-session RNN itself forms a baseline for the II-RNN.

\paragraph{Most Popular}
The most popular baseline is a very simple baseline, but it can perform decently in some cases. All items are sorted by their number of occurrences in the training set, and the top k items are recommended at each time step. Although a very basic baseline, it provides a nice sanity check. Any serious model should be able to beat this model.

\paragraph{Most recent}
Even though we removed consecutive repetitions of items in all sessions, there could still be a high repetitiveness of items within sessions (i.e. some items can occur multiple times in a session). Especially in the Reddit and Last.fm datasets, where users can tend to interact with some subreddits or artists multiple times in their sessions. We believe that it is less likely to see such repetitiveness in the Instacart dataset, because users probably only add each item to their cart once. The most recent baseline behaves as a stack. It is initially filled with k random items. For each time step, the item interacted with is added to the top of the stack, and the item at the bottom is pushed out of it. However, if the new item is already in the stack, it is just moved to the top. The recommendation at each time step is then the stack of recently seen items, where the top recommendation is the item just interacted with. Our model should be able to beat this baseline significantly. But the most recent baseline gives us information about the diversity of items within sessions

\paragraph{Item-kNN}
Item-k nearest neighbors (Item-kNN) is a simple, but usually strong baseline. It is commonly used in practice as a item-to-item recommender \cite{Linden2003}. Different implementations are possible. We implemented it as follows. For each item in the dataset, we count the number of co-occurrences with the other items in the dataset. A co-occurrence is when two items appear in the same session. When testing, the algorithm recommends the top $k$ items with highest co-occurrences with the last seen item.

\paragraph{BPR-MF}
The Bayesian Personalized Ranking for Matrix Factorization (BPR-MF) \cite{Rendle2009} is a commonly used matrix factorization method. It tries to predict personal pairwise rankings of unseen items (i.e. given a user and two items, BPR- MF tries to predict which of the two items the user would rate higher). We use an existing implementation\footnote{theano-bpr: \url{https://github.com/bbc/theano-bpr}}, that we tweak slightly to fit our use case. The original implementation does not recommend already seen items, but in our case, users often interact with items they have already seen.
BPR-MF computes feature vectors for users and items based on the users earlier interactions, and is then able to make a recommendation based on this. This means that the recommendations will be the same throughout future sessions, unless the model is re-trained. In other words, BPR-MF cannot be applied directly to session-based recommendations. To make a more fair comparison, we create a new split of the datasets. Only the last session of each user is put in the test set. BPR-MF still produce the same recommendations for all time steps in the test session for a given user.

\subsection{Evaluation and Hyperparameters Tuning}
We used \textit{Recall@$K$} and \textit{MRR@$K$} with $K = 5,10,20$ to evaluate all models. In addition to the baselines already discussed, we also compared the intra-session RNN to the II-RNN on the two presented datasets. We experimented with mini-batch sizes, embedding sizes, learning rate, dropout rate, using multiple GRU layers, and number of session representations to find the best configurations for each dataset. 
The best configurations we found are summarized in Table \ref{tab:hyperparam}. We employed two different configurations of the II-RNN, one using average-pooling and the other using last hidden state as session representations for past sessions. We used the same size for the item embeddings and internal vectors in the GRU layers. We found $tanh$ to work well as activation function in the GRU layers, and did not investigate other alternatives.

\begin{table}
\begin{center}
  \begin{tabular}{lcc}
    \toprule
     &Reddit&Last.fm\\
    \midrule
    Embedding size & 50& 100\\
    Learning rate & 0.001& 0.001\\
    Max. recent session representations & 0& 0.2\\
    Mini-batch size & 15& 15\\
    Number of GRU layers, intra-session level & 100& 100\\
    Number of GRU layers, inter-session level & 1& 1\\
  \bottomrule
\end{tabular}
\end{center}
\caption{Best configurations for the RNN models. We found that the configurations that worked well for the II-RNN, worked well for the standalone intra-session RNN as well. Not all configurations are applicable to the standalone intra-session RNN.}
\label{tab:hyperparam}
\end{table}

\subsection{Creating mini-batches}
We want our model to be biased towards recent user trends. This is often desirable in practice, and we find it reasonable to assume that it applies for our datasets. Furthermore, the way we split our dataset into training- and test sets reflect this. I.e. the test set contains the most recent samples for each user. This leaves us with two desirable properties for how the training samples should be processed. First, more recent samples should be processed last. Second, each mini-batch should contain a variety of users. I.e. no user should be over represented with samples in any mini-batch.
To achieve these properties, we constructed the following scheme for creating mini-batches. Each training sample, a session, is associated with a user. All sessions belonging to the same user, are grouped together and sorted oldest to newest.

\subsection{Implementation Details}
The implementation is done in Python 3.5.2, with the Tensorflow machine learning software library. We run our experiments on three different computers, all with the Ubuntu 16.04 operating system. All computers have at least 16 GB of RAM, and a Nvidia GeForce GTX 1060 6 GB or better. The code is available on github here\footnote{\url{https://github.com/olesls/master_thesis}}.

\section{Results and Discussion}
\label{subsec:results}
We evaluate the performance of the proposed models by using the standard evaluation metrics presented in the previous Section. The comparison is performed over the baselines from literature over the two datasets, Last.fm and Reddit.

\subsection{Inter-session model effectiveness}
We found that using multiple GRU layers did not improve performance neither when applied at the inter-session level, nor at the intra-session level. Dropout was crucial in order to get good results on the Last.fm dataset, while on the Reddit dataset the models got better results without dropout. To achieve the best results, dropout had to be used on all GRU layers.

Table \ref{tab:redditres} shows an overview of how the models and baselines scored on the \textit{Reddit} dataset. Relative scores are given compared to the standalone intra-session RNN, considered the strongest baseline. We ran the RNN model three times and the results presented in the table are averages of three runs, even though the results were usually consistent between runs. 
Similarly, Table \ref{tab:lastfmres} presents the results for the \textit{Last.fm} dataset. 
For both datasets, Item-kNN and RNN were the strongest baselines, but were both clearly outperformed by the intra-session RNN. 

When it comes to the two versions of II-RNN (AP using average pooling and LHS using the last hidden state, 
cf.\ Figures \ref{fig:iiRNNavgpooling-used} and \ref{fig:iiRNNlasthidden}), Table \ref{tab:redditres} shows that using the last hidden state of the intra-session RNN as the representation of a session is slightly better than using average pooling for the Reddit dataset, while the results for the Last.fm dataset (Table \ref{tab:lastfmres})
are reversed, this time with AP being better than LHS. 


\begin{table*}[ht!]
\begin{subtable}{1\textwidth}
\centering
\scalebox{.8}{
  \begin{tabular}{rcccccc}
    \toprule
     &R@5&R@10&R@20&MRR@5&MRR@10&MRR@20\\
    \midrule
    Item-KNN & 0.2171& 0.3032& 0.3885& 0.1174& 0.1288& 0.1349\\
    Most Recent & 0.2152& 0.2205& 0.2209& 0.0969& 0.0977& 0.0977\\
    Most Popular & 0.1322& 0.1946& 0.2647& 0.0850& 0.0932& 0.0982\\
    RNN & 0.3372& 0.4173& 0.5004& 0.2436& 0.2542 & 0.2600\\
    II-RNN-AP & \makecell{0.4361 \\ (+29.3\%)} & \makecell{0.5168 \\ (+23.8\%)} & \makecell{ 0.5963 \\ (+19.2\%)} & \makecell{0.3202 \\ (+31.4\%)} & \makecell{0.3309 \\ (+30.1\%)} & \makecell{0.3364 \\ (+29.4\%)} \\
    \bf{II-RNN-LHS} & \makecell{\bf{0.4476} \\ \bf{(+32.7\%)}} & \makecell{\bf{0.5344} \\ \bf{(+28.1\%)}} & \makecell{\bf{0.6180} \\ \bf{(+23.5\%)}} & \makecell{\bf{0.3213} \\ \bf{(+31.9\%)}} & \makecell{\bf{0.3329} \\ \bf{(+31.0\%)}} & \makecell{\bf{0.3388} \\ \bf{(+30.3\%)}} \\
  \bottomrule
\end{tabular}
}
\caption{Reddit Dataset.}
\label{tab:redditres}
\end{subtable}

\begin{subtable}{1\textwidth}
\centering
\scalebox{.85}{
  \begin{tabular}{rcccccc}
    \toprule
     &R@5&R@10&R@20&MRR@5&MRR@10&MRR@20\\
    \midrule
    Item-KNN & 0.0851& 0.1191& 0.1590& 0.0504& 0.0548& 0.0576\\
    Most Recent & 0.1061& 0.1305& 0.1379& 0.0422& 0.0456& 0.0462\\
    Most Popular & 0.0528& 0.0650& 0.0829& 0.0433& 0.0449& 0.0462\\
    RNN & 0.1350& 0.1843& 0.2478& 0.0867& 0.0932 & 0.0976\\
    \bf{II-RNN-AP} & \makecell{\bf{0.1478} \\ \bf{(+9.5\%)}} & \makecell{\bf{0.2048} \\ \bf{(+11.1\%)}} & \makecell{\bf{0.2788} \\ \bf{(+12.5\%)}} & \makecell{\bf{0.0930} \\ \bf{(+7.3\%)}} & \makecell{\bf{0.1005} \\ \bf{(+7.8\%)}} & \makecell{\bf{0.1056} \\ \bf{(+8.2\%)}} \\
    II-RNN-LHS & \makecell{0.1439 \\ (+6.6\%)} & \makecell{0.2018 \\ (+9.5\%)} & \makecell{0.2776 \\ (+12.0\%)} & \makecell{0.0891 \\ (+2.8\%)} & \makecell{0.0968 \\ (+3.9\%)} & \makecell{0.1020 \\ (+4.5\%)} \\
  \bottomrule
\end{tabular}
}
\caption{Last.fm Dataset.}
\label{tab:lastfmres}
\end{subtable}

\caption{Recall and MRR scores for the II-RNN models and the baselines. Relative scores are given compared to the standalone intra-session RNN. The best results per dataset are highlighted. The two II-RNN models differ by how they feed information to the inter-session model, either using average pooling (II-RNN-AP) or the last hidden state (II-RNN-LHS).}
\end{table*}

Finally, Table \ref{tab:redditresBPRFM} and \ref{tab:lastfmresBPRFM} shows how the BPR-MF baseline performed on the hold-one-out version of the dataset. Due to limited space, we only show the results for the BPR-MF. In all cases, the II-RNN significantly outperformed the standalone intra-session RNN and the BPR-MF method. 

\begin{table*}[ht!]
\begin{subtable}{1.0\textwidth}
\centering
\scalebox{.9}{
  \begin{tabular}{rcccccc}
    \toprule
     &R@5&R@10&R@20&MRR@5&MRR@10&MRR@20\\
    \midrule
    BPR-MF & 0.1271& 0.1900& 0.2621& 0.0878& 0.0961& 0.1011\\
    RNN & 0.3660& 0.4388& 0.5118& 0.2781& 0.2878 & 0.2928\\
    \bf{II-RNN-LHS} & \bf{0.5022}  & \bf{0.5803}  & \bf{0.6537} & \bf{0.3807} &  \bf{0.3912}  & \bf{0.3963} \\
  \bottomrule
\end{tabular}
}
\caption{Reddit Dataset.}
\label{tab:redditresBPRFM}
\end{subtable}

\begin{subtable}{1.0\textwidth}
\centering
 \scalebox{.9}{
 \begin{tabular}{rcccccc}
    \toprule
     &R@5&R@10&R@20&MRR@5&MRR@10&MRR@20\\
    \midrule
    BPR-MF & 0.0619& 0.0833& 0.1207& 0.0467& 0.0494& 0.0520\\
    RNN & 0.1568& 0.2088& 0.2761& 0.0972& 0.1041 & 0.1088\\
    \bf{II-RNN-AP} & \bf{0.1775} & \bf{0.2390}  & \bf{0.3133}  & \bf{0.1085} & \bf{0.1165}  & \bf{0.1216}  \\
  \bottomrule
\end{tabular}
}
\caption{Last.fm Dataset.}
\label{tab:lastfmresBPRFM}
\end{subtable}
\caption{Recall and MRR scores for the BPR-MF baseline and the RNN models on the hold-one-out version of the dataset. Only the best performing II-RNN model is included for each dataset.}
\end{table*}

\subsection{Impact on Session Cold start problem}
\begin{figure*}[ht!]
\centering
	\begin{subfigure}[t]{3in}
		\centering
        \includegraphics[width=1.12\textwidth]{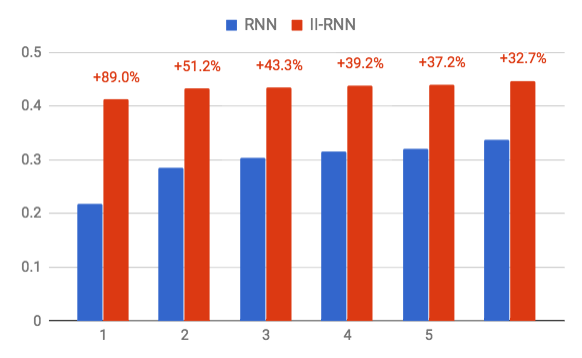}
		\caption{Reddit Dataset.}\label{fig:iiRNN-coldstart-reddit}
	\end{subfigure}
    \hspace{0.06\textwidth}
    \begin{subfigure}[t]{3in}
		\centering
        \includegraphics[width=1.12\textwidth]{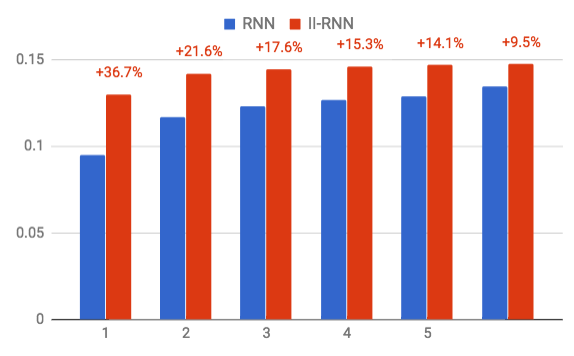}
		\caption{Last.fm Dataset.}\label{fig:iiRNN-coldstart-lastfm}
	\end{subfigure}
 	\caption{Effect of cold-start at the offset of a new session.}
\end{figure*}

The intra-session RNN learns about the user as it observes item interactions throughout a session. It is therefore reasonable to believe that the model's prediction accuracy increases throughout the session. As discussed, the II-RNN can improve both the overall recommendations, and especially the first few recommendations in each session, impacting strongly on the cold start problem within a session. To evaluate this, we test the RNN models both on the overall recommendations and on the first $n$ recommendations in a session, for $n = 1,\ldots, 5, L$, where $L$ is the maximum session length. 
That is, we evaluate the models on recommendations for the first $n$ time steps, and note that when $n = L$ we retain the overall score already reported.
A comparison between RNN and II-RNN-LHS in terms of \textit{Recall@}5 are shown in Figure \ref{fig:iiRNN-coldstart-reddit} for the Reddit dataset. Notice how the II-RNN already at the first recommendation of a new session achieves $R@5 > .4$, a substantial $89\%$ improvement over the RNN model. While the RNN catches up somewhat as more interactions are seen in the current session, the II-RNN also improves with more information, and holds a $32.7\%$ improvement over the RNN at the end of the session. 
Similar results can be seen in Figure \ref{fig:iiRNN-coldstart-lastfm}, where II-RNN-AP is compared to the RNN using the Last.fm dataset. Again, we see a dramatic improvement early on in a new session, and even though the RNN catches up some of the II-RNN's $36.7\%$ lead, the II-RNN remains superior throughout the session. 

\section{Conclusion}
\label{sec:conclusion}
In this paper we have investigated a new RNN architecture for session-based recommendations, termed II-RNN. II-RNN combines modeling of recommendations inside a single session with an inter-session RNN that serves as a memory of user interactions from historical sessions. The two parts are combined into a single architecture. We have evaluated II-RNN using two publicly available datasets, and show considerable improvements over strong baselines. Furthermore, we found the II-RNN model to be particularly adept to making recommendations early in a user session, thereby helping to alleviate the well-known \textit{cold-start} problem session-based recommender systems are confronted with.

We anticipate at least two paths for future research: Firstly, while the II-RNN model already works well 
using either of the two methods for creating session representations (AP and LHS), we will consider other approaches as well. Further improvement can potentially be achieved by for example considering more complex methods for representing each session, or by using other more advanced attention mechanisms. Secondly, we are currently  utilizing time-information only  implicitly through the notion of sessions. We believe that explicitly representing the time-difference between sessions will improve the recommendations, and are currently investigating how to efficiently incorporate this information into the recommendation process.

\bibliographystyle{ACM-Reference-Format}
\bibliography{sigproc}

\end{document}